\documentclass[preprint,preprintnumbers,amsmath,amssymb,nofootinbib]{revtex4}
 
\usepackage{graphicx}
\usepackage{dcolumn}
\usepackage{bm}
\usepackage{verbatim}

\begin{document}

\title{A PNJL model in 0+1 Dimensions}
\author{K. Dusling}  
\affiliation{Department of Physics, Brookhaven National Laboratory, Upton, New York 11973-5000, USA}
\affiliation{Department of Physics \& Astronomy, State University of
 New York, Stony Brook, NY 11794-3800, U.S.A.}
\author{C. Ratti, I. Zahed}
\affiliation{Department of Physics \& Astronomy, State University of
 New York, Stony Brook, NY 11794-3800, U.S.A.}
\date{\today}  

\begin{abstract}  

We formulate the Polyakov-Nambu-Jona-Lasinio (PNJL) model in 0+1 dimensions. The
thermodynamics captured by the partition function yields a bulk pressure, as well
as quark susceptibilities versus temperature that are similar to the ones in 3+1 
dimensions. Around the transition temperature the behavior in the pressure and
quark susceptibilities follows from the interplay between the lowest Matsubara
frequency and the Polyakov line. The reduction to the lowest Matsubara frequency
yields a matrix Model. In the presence of the Polyakov line the UV part of the Dirac spectrum features oscillations when close to the transition temperature. 

\end{abstract} 


\maketitle  

\section{Introduction}

There has been a large success in modeling the finite temperature behavior of QCD using the Nambu--Jona-Lasinio (NJL) 
model~\cite{Nambu:1961tp,Nambu:1961fr}.  The NJL model is based on an effective Lagrangian of relativistic quarks 
interacting through a local and chirally symmetric four-point interaction.  It was suggested that this model may 
serve as a good approximation to the low-lying chiral excitations of the QCD vacuum as well as the QCD thermodynamics
below the transition temperature $T_c$. Key in the NJL model is the spontaneous breaking of chiral symmetry and the
emergence of a chiral constituent quark mass, which is generated through the interaction of quarks
with the chiral condensate. 

The main drawback of the NJL model is that it does not include the properties of color confinement.  This leads to 
the problem that the model contains the wrong degrees of freedom near the transition temperature $T_c$.  This has 
led to the development of extended NJL models which include some effects of confinement by introducing the Polyakov 
loop as a new classical field which couples to quarks.  These models are referred to as Polyakov-loop-extended NJL 
(PNJL) models \cite{Meisinger:1995ih,Fukushima:2003fw,agnes,Megias:2004hj,Ratti:2005jh,Roessner:2006xn}. Many aspects of these models have been extensively 
investigated recently, including thermodynamics and phase structure for two
\cite{thermodynamics}, and three \cite{threeflavor} flavor systems, finite 
isospin systems \cite{finite_isospin}, imaginary chemical potential
\cite{imaginary}, mesonic
modes \cite{mesons} and studies related to the fermionic sign problem and
incorporation of fluctuations \cite{more_theoretical}.

In this work we recast the PNJL model into a simple effective Lagrangian in 0+1 dimensions.  
Modifications to the thermodynamics and susceptibilities as compared to the four dimensional case are discussed.  
We show that the key features of the four dimensional physics across the transition temperature are captured
by the interplay of one Matsubara frequency against the Polyakov line. The model with one Matsubara frequency
reduces to a matrix model. The resulting Dirac spectrum oscillates near $T_c$. This paper is
organized as follows: in section 2 we formulate
the model. In section 3 we derive the phase diagram, the bulk pressure and quark susceptibilities. In section 
4 we detail the matrix model and derive the pertinent mean-field equations for the resolvent. The Dirac spectra
are constructed numerically for temperatures across $T_c$. Our conclusions are in section 5.
 
\section{Model} 

Motivated by the recent work on the PNJL model 
\cite{Fukushima:2003fw,agnes,Megias:2004hj,Ratti:2005jh,Roessner:2006xn,thermodynamics,threeflavor,finite_isospin,imaginary,mesons,more_theoretical} we consider the following 
schematic Lagrangian density in one-dimension, including a NJL four-fermion 
contact term and coupling to a 
constant temporal background gauge field, whose dynamics is encoded in the 
phenomenological 
potential $U$:
\begin{equation}
\mathcal{L}_{4}=\psi^\dagger(i\gamma_4 D^4+im+i\mu\gamma_4)\psi+
\frac{g^2}{2}\biglb( (\psi^\dagger\psi)^2+(\psi^\dagger i\gamma_5 \psi)^2\bigrb)+U(\phi[A],\phi^*[A],T)
\label{1}
\end{equation}
In the above equation, $\psi_{a,f}$ are quark fields where $a=1,2,...,N_c$ are color indices and $f=1,2,...,N_f$ are 
flavor indices.  For simplicity $N_f=1$ unless specified otherwise.  $D^4=\partial_\tau+iA^4$ is the covariant 
deriviative, $m$ is the bare quark mass and $A^4=G\mathcal{A}^4_a{\lambda_a/2}$ where 
$\mathcal{A}^4_a$ is the temporal component of the SU(3) 
gauge field, G is the gauge coupling, and $\lambda_a$ are the Gell-Mann matrices.
We consider scalar fermions in our work, therefore the $\gamma$ matrices in Eq. ({1}) are 2$\times$2
matrices.
 The axial-anomaly and the effects
of $U(1)_A$ breaking will be discussed elsewhere.

The mean-field analysis of (\ref{1}) is readily carried out by the bosonization procedure which consists in replacing
the four-quark interaction with color-singlet auxiliary fields defined as
\begin{eqnarray}
P=-2ig^2\langle\psi^\dagger_L\psi_L\rangle \nonumber\,,\\
P^\dagger=-2ig^2\langle\psi^\dagger_R\psi_R\rangle\,,
\end{eqnarray}
so that
\begin{eqnarray}
\mathcal{L}_{4}&=&\psi^\dagger_Ri(\partial_4+\mu+iA^4)\psi_L+\psi^\dagger_Li(\partial_4+\mu+iA^4)\psi_R  \nonumber\\
&&+\psi^\dagger_Li(P^\dagger+m)\psi_L+\psi^\dagger_Ri(P+m)\psi_R+\frac{1}{2g^2}PP^\dagger+U(\phi[A],\phi^*[A],T)\,,
\label{eq:3}
\end{eqnarray}
in the chiral basis, $\psi=(\psi_R,\psi_L)$.  The effective potential for the background gauge field is expressed 
in terms of the traced Polyakov loop.  We work in the Polyakov gauge and take the gauge field $A^4$ as time-independent.  
The traced Polyakov loop is then expressed as
\begin{equation}
\phi[A]=\frac{1}{N_c}\text{ Tr}_c L \,.\nonumber\\
\end{equation}
In our gauge choice, the Polyakov loop matrix is diagonal and defined as
\begin{equation}
L=\exp\biglb[\frac{iA^4}{T}\bigrb]=\text{diag}(e^{i\nu_1},e^{i\nu_2},e^{-i(\nu_1+\nu_2)})\,.
\end{equation}
While $L$ is gauge dependent, $\phi[A]$ is gauge invariant and in general complex valued. The 
potential for the Polyakov line
$U(\phi , \overline\phi, T)$ satisfies the Z(3) center symmetry. At low temperatures we expect the potential to have a minimum 
at $\phi=0$.  At temperatures above $T_0$ it develops a minimum which gradually forces $\phi\to 1$ as 
$T\to\infty$.  The potential in \cite{Ratti:2005jh}, which is used here as well, was fit in order to reproduce the pure-gauge 
lattice data in 3+1 dimensions:  
\begin{equation}
\frac{U(\phi,\bar{\phi},T)}{T}=\frac{-b_2(T)}{2}\bar{\phi}\phi-\frac{b_3}{6}(\phi^3+\bar{\phi}^3)+\frac{b_4}{4}(\bar{\phi}\phi)^2\,,
\end{equation}
where
\begin{equation}
b_2(T) = a_0+a_1\left(\frac{T_0}{T}\right)+a_2\left(\frac{T_0}{T}\right)^2+a_3\left(\frac{T_0}{T}\right)^3\,.
\end{equation}
The coefficients were fit in \cite{Ratti:2005jh} to the lattice data for pure gauge QCD thermodynamics.  They are given 
as $a_0=6.75, a_1=-1.95, a_2=2.625, a_3=-7.44, b_3=0.75,\text{ and } b_4=7.5$.
 
The partition function corresponding to the above action is given as 
\begin{equation}
Z=\int [d\psi][d\psi^\dagger][d\phi][d\bar{\phi}][dP]e^{-S}\,,
\end{equation}
where
\begin{equation}
S=\int_{\frac{1}{T}\times V_3} \mathcal{L}_4 \text{ }d^4x=V_3\int_\beta \mathcal{L}_4 \text{ }d\tau\,.
\end{equation}
Making use of the anti-periodicity of the quark-fields $\psi(\tau+\beta)=-\psi(\tau)$ and the fact that at finite 
temperature the operator $i\partial_4$ is invertible with a discrete spectrum $\omega_n=(2n+1)\pi T$ the path 
integration over the fermionic fields can be done resulting in the following form of the partition function
\begin{equation}
Z=\int[dP][d\phi]e^{-V_3\int_\beta [\frac{1}{2g^2}PP^\dagger+U(\phi,T)]}\prod_{n=-\infty}^{\infty}\prod_{a=1}^{N_c}\text{det}_2
\left( \begin{array}{cc}i(m+P) & \omega_n+i\mu-A^4 \\ \omega_n+i\mu-A^4 & i(m+P^\dagger) \end{array} \right)\,.
\label{eq:pf}
\end{equation}

\section{Thermodynamics}

\begin{figure}
\includegraphics[scale=1]{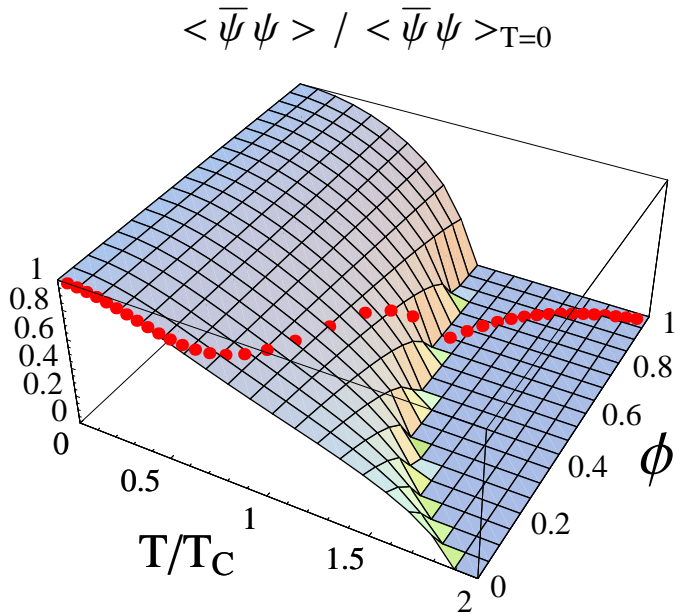}
\caption{Scaled chiral condensate, $\langle \overline{\psi}\psi\rangle / \langle \overline{\psi}\psi\rangle_{T=0}$, as a function of $T$ and $\phi$.  The dotted red line shows a schematic 
         trajectory when $\phi(T)$ is taken from the mean field calculation.}
\label{fig1}
\end{figure}

We now discuss the thermal properties of the above model in the mean field (saddle point) approximation.  
The thermodynamic potential associated with equation (\ref{eq:pf}) is
\begin{equation}
\Omega=U(\phi,\bar{\phi},T)+\Sigma N_c P^2-T\sum_{n=-\infty}^{\infty} \text{Tr}_c\ln[\beta^2 \omega^2+\beta^2(\omega_n+i(\mu+iA_4))^2]\,, 
\label{eq:epsum}
\end{equation}
where $\Sigma=1/(2g^2)$\footnote{The value of $\Sigma$ is chosen such that the constituent quark mass at zero temperature is $P(T=\mu=0)=\frac{1}{2\Sigma}=$ 300 MeV, a value consistent with NJL models in 3+1 dimensions.}  The above series does not converge and a divergent, temperature-independent piece needs to be subtracted (see for example \cite{kapusta}).
Following this procedure, the sum over $n$ can be done explicitly, which up to an overall constant yields
\begin{equation}
\Omega=U(\phi,\bar{\phi},T)+\Sigma N_c P^2-N_c\omega-T\text{ Tr}_c\ln[1+L e^{-\beta(\omega-\mu)}] 
-T\text{ Tr}_c\ln[1+L^\dagger e^{-\beta(\omega+\mu)}]\,,
\end{equation}
where $\omega=|P+m|$.  Using the identity $\text{Tr} \ln(X) = \ln \text{det}(X)$ we arrive at the following form for the effective potential:
\begin{eqnarray}
\Omega=U(\phi,\bar{\phi},T)+\Sigma N_c P^2-N_c\omega&-&T\ln[1+3\phi e^{-\beta(\omega-\mu)}
+3\bar{\phi} e^{-2\beta(\omega-\mu)}+e^{-3\beta(\omega-\mu)}] 
\nonumber\\&-&T\ln[1+3\bar{\phi} e^{-\beta(\omega+\mu)}+3\phi e^{-2\beta(\omega+\mu)}+e^{-3\beta(\omega+\mu)}]. 
\label{eq:epfinal}
\end{eqnarray}
In the saddle-point approximation, the values of $P, \phi$ and $\bar{\phi}$ that maximize the above potential are 
found by numerically solving the coupled system of equations:
\begin{align}
\frac{\partial\Omega}{\partial P}=0\text{,     }\frac{\partial\Omega}{\partial \phi}=0
\text{,     }\frac{\partial\Omega}{\partial \bar{\phi}}=0\,.
\end{align}
The solution of these equations then provides mean field values which can be used in evaluating any thermodynamic quantities, such 
as the pressure $p(T,\mu)=-\Omega(T,\mu)$.  The goal is two-fold.  First, one would like to see if the above {\it simplified}
PNJL model in 0+1 dimensions can reproduce the bulk properties of the PNJL model in four dimensions.  Secondly, one would like to 
further reduce this model to a matrix model in zero dimensions by including only a finite number of matsubara frequencies in the 
sum of equation~(\ref{eq:epsum}).  
The chiral condensate as a function of temperature and traced Polyakov loop is plotted in figure~\ref{fig1}.  The dotted curve 
shows the mean-field trajectory for $\phi$.

\begin{figure}[hbtp]
  \vspace{9pt}
  \centerline{\hbox{ \hspace{0.0in} 
\includegraphics[scale=.7]{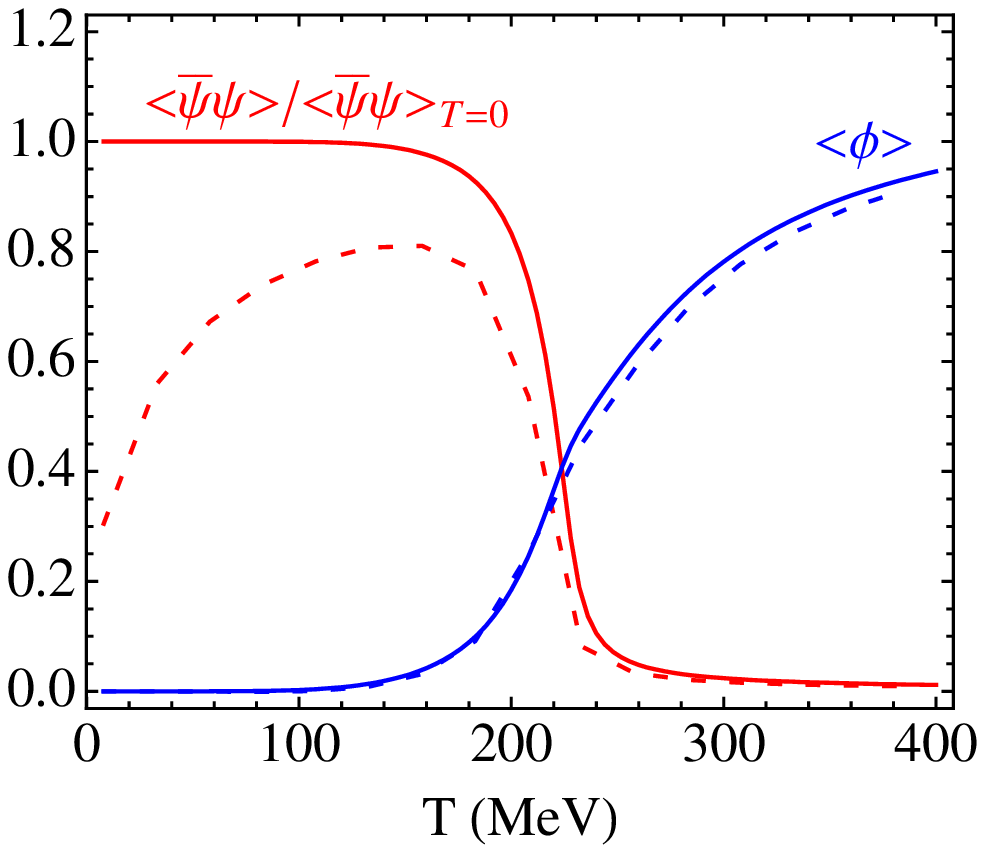}
    \hspace{0.1in}
\includegraphics[scale=.7]{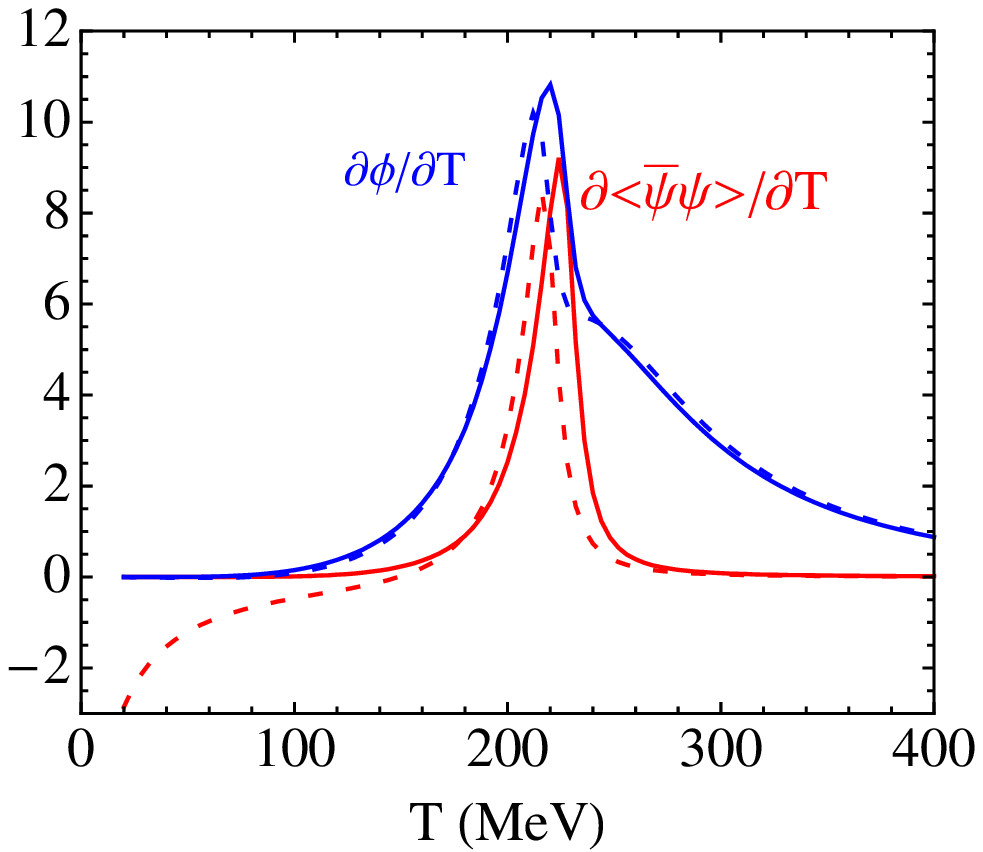}
    }
  }
   \vspace{9pt}
  \caption{(Color online) Left:  Scaled chiral condensate $\langle \bar{\psi}\psi\rangle / \langle \bar{\psi}\psi\rangle_{T=0}$ (red curve) and Polyakov loop (blue curve).  Right: $\partial\langle \bar{\psi}\psi\rangle /\partial T$ (red) and $\partial \phi / \partial T$ (blue).  In both figures, the solid curves are evaluated by using the sum over all matsubara frequencies and the dashed curves by using the two lowest frequencies, $\pm \pi T$.}
 \label{fig:cond}
\end{figure}

In the left panel of figure~\ref{fig:cond} we show both the chiral condensate and Polyakov loop as functions of temperature at $\mu=0$.  
The solid lines are the results from the model in $0+1$ dimensions, obtained by minimizing
the thermodynamic potential of eq. (\ref{eq:epfinal}). Figure 2 in reference \cite{Ratti:2006gh} shows 
the same quantities for the model in four dimensions.  We find qualitatively the same behavior for both the condensate, 
Polyakov loop, as well as the two susceptibilities ($\partial\langle \bar{\psi}\psi\rangle /    \partial T$ 
and $\partial \phi / \partial T$). Also shown in these figures are the results using only the two lowest matsubara 
modes in equation~(\ref{eq:epsum}) as dashed curves. For temperatures $T>100$ MeV the sum 
over the first two frequencies is a good approximation to the infinite sum. The vanishing of the chiral condensate
at low temperature for the truncated frequencies is due to the occurence of $\beta=1/T$ in the weight factor in
(\ref{eq:pf}), which vanishes as $\beta=1/T\rightarrow \infty$. When {\it all} Matsubara modes are included, this 
vanishingly small weight factor is overcome by the determinant part with infinitly many modes, leading to
a finite chiral condensate at zero temperature as it should. We will come back to this point in the random matrix
reduction.

\begin{figure}[!ht] 
\includegraphics[scale=1.1]{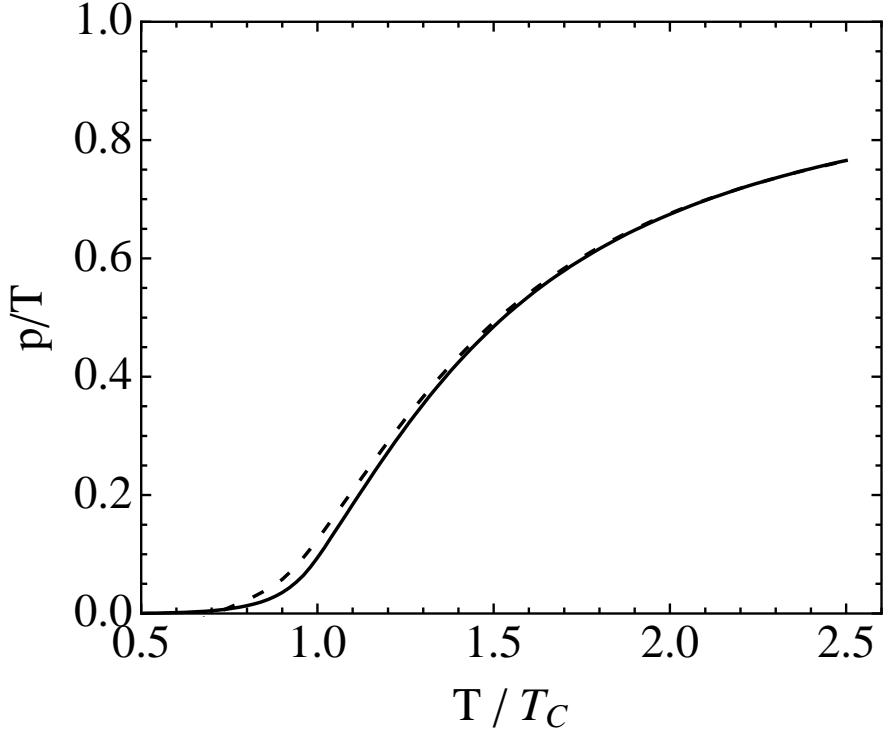}
\caption{Scaled pressure normalized to the ideal gas result.  The solid line is from eq.~(\ref{eq:epfinal}) 
while the dashed line is the result keeping only the two lowest matsubara frequencies, $\pm \pi T$.}
\label{fig:PovT}
\end{figure}

\begin{figure}
\includegraphics[scale=1.1]{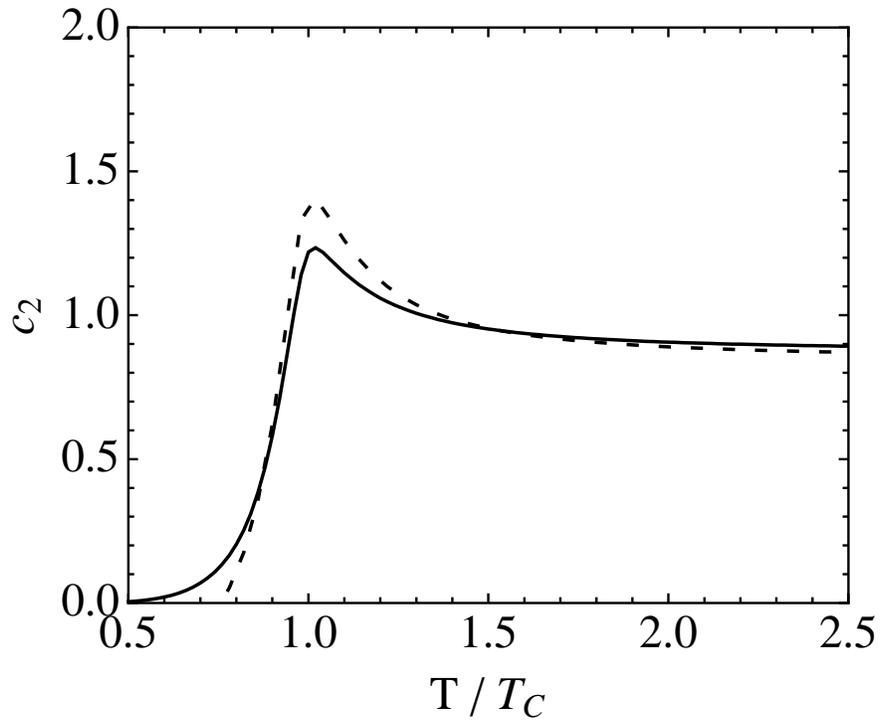}
\caption{$c_2$ as a function of $T/T_c$ normalized to the ideal gas value.}
\label{fig:c2}
\end{figure}

\begin{figure}
\includegraphics[scale=1.1]{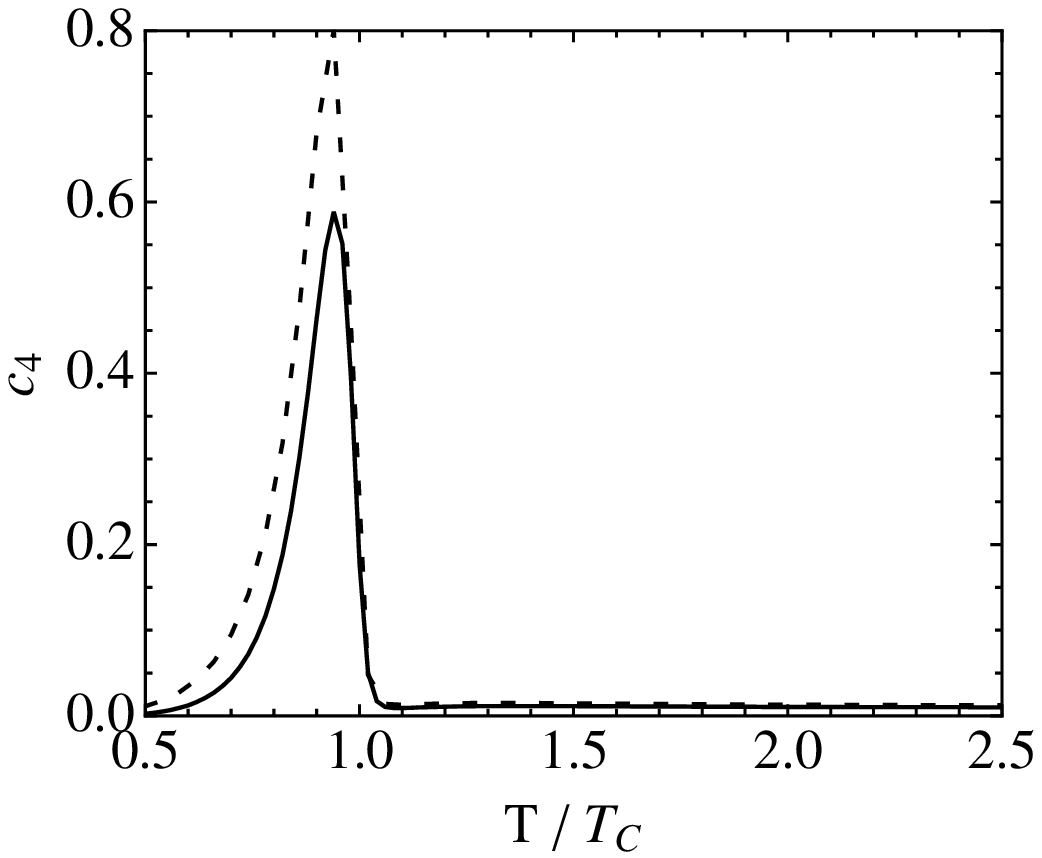}
\caption{$c_4$ as a function of $T/T_c$.}
\label{fig:c4}
\end{figure}

\begin{figure}
\includegraphics[scale=1.1]{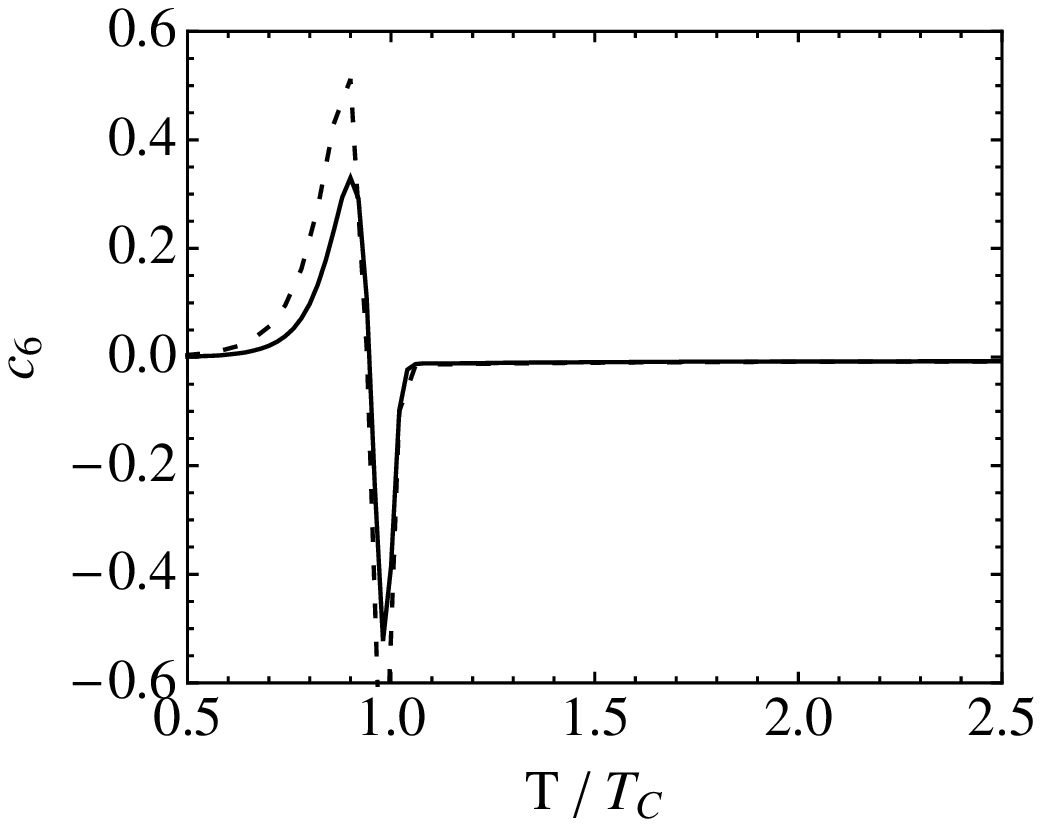}
\caption{$c_6$ as a function of $T/T_c$.}
\label{fig:c6}
\end{figure}

\begin{figure}
\includegraphics[scale=1.1]{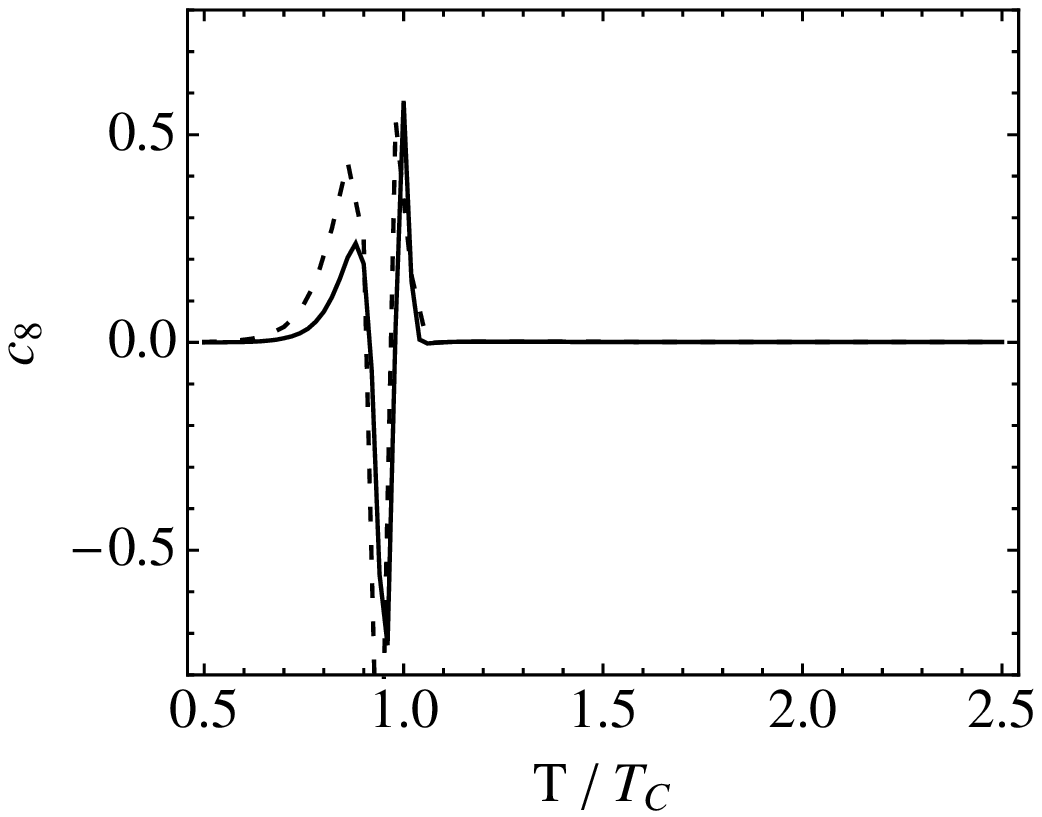}
\caption{$c_8$ as a function of $T/T_c$.}
\label{fig:c8}
\end{figure}

We now show the results for the quark number susceptibilites using both 
eq.~(\ref{eq:epfinal}), where the explicit sum has 
been carried out over all matsubara frequencies, and eq.~(\ref{eq:epsum}) where
the sum includes only the two lowest 
frequencies ($\pm \pi T$).  The coefficients are extracted up to eighth order 
by a fit to the scaled pressure
\begin{equation}
\frac{p(T,\mu)}{T}=\sum_{n=0}^\infty c_n(T)\biglb( \frac{\mu}{T} \bigrb)^n\,.
\end{equation}
First we should compare our results with those of reference~\cite{Ghosh:2006qh} where this exercise was carried out 
in 3+1 dimensions.  The solid curves in figures \ref{fig:PovT}-\ref{fig:c8} show the scaled pressure and first four 
susceptibilities for the PNJL model in 0+1 dimensions where the sum is performed over all frequencies.  Qualitatively, 
similar behavior is obtained in both the 0+1 dimensional model used in this work and the four dimensional model used 
in \cite{Ghosh:2006qh}.  Near the transition temperature $T_c$ the peak structures are again similar in both models as 
seen in $c_2$ through $c_8$, which shows the direct interplay between the Polyakov line and the lowest Matsubara
frequencies.  There are qualitative differences in the high temperature behavior.  This is due to differences in the 
dimensionality of the problem which we now discuss.    

In order to understand the effect of fewer dimensions and see if there are any qualitative differences between the 
model in four and one dimension we look at the high temperature limit (no longer mean field) of an ideal gas of quarks 
and anti-quarks.  In four dimensions the pressure is given as
\begin{equation}
\frac{p}{T^4}=\frac{N_c N_f}{\pi^2 T^3}\int_0^\infty dk\text{ } k^2 \ln\biglb[ (1-n)(1-\bar{n}) \bigrb]\,,
\end{equation}
where $n=1/(1+e^{-(\sqrt{k^2+m^2}+\mu)/T})$ and $\bar{n}=1/(1+e^{-(\sqrt{k^2+m^2}-\mu)/T})$  which leads to a finite 
series in chemical potential for massless quarks

\begin{equation}
\frac{p}{N_c N_f T^4}=\frac{7\pi^2}{180}+\frac{1}{6}\biglb( \frac{\mu}{T} \bigrb)^2+\frac{1}{12\pi^2}\biglb( \frac{\mu}{T} \bigrb)^4\,.
\end{equation}
Immediately one can extract the high temperature behavior for the susceptibilities in four dimensions: 
$c_2\to\frac{N_c N_f}{6},$ $c_4\to\frac{N_c N_f}{12\pi^2}$ and $c_6=c_8=0$.  However, in the 0+1 dimensional NJL model, 
the pressure is given as
\begin{equation}
\frac{p}{T}=N_c N_f \ln\biglb[ (1-n)(1-\bar{n}) \bigrb]\,,
\end{equation}
which for massless quarks at high temperature ({\em i.e.} $n,\bar{n}=1/[1+e^{\mp\mu/T}]$)  leads to the following result
\begin{equation}
\frac{p}{N_c N_f T}=\ln 4+\frac{1}{4}\biglb( \frac{\mu}{T} \bigrb)^2-\frac{1}{96}\biglb( \frac{\mu}{T} \bigrb)^4
+\frac{1}{1440}\biglb( \frac{\mu}{T} \bigrb)^6-\frac{17}{322560}\biglb( \frac{\mu}{T} \bigrb)^8+\cdots\,.
\end{equation}
This leads to different asymptotic susceptibilities in 0+1 dimensions: $c_2\to\frac{N_c N_f}{4},$ and now 
$c_4\to-\frac{N_c N_f}{96}$ has changed sign.  Also, $c_6$ and $c_8$ are non-vanishing.  Therefore, at least in the high 
temperature limit, one should expect qualitative differences between the model in four and one dimensions.  

The dashed curves in figures \ref{fig:PovT}-\ref{fig:c8} show the same result using only the two lowest Matsubara 
frequencies in the energy sum.  At temperatures close to $T_c$ the qualitative structure of the susceptibilities is 
reproduced.  At higher temperatures the finite sum result approaches the full result.  We note that in the current 
model the high temperature limit of the susceptibilities is never reached.  This is due to the fact that $\phi$ 
and $\bar{\phi}$ are treated as independent variables and, as pointed out in \cite{Roessner:2006xn}, this tends 
to overestimate the difference between $\phi$ and $\bar{\phi}$.  If we set $\phi=\bar{\phi}$, the ideal gas result 
would be obtained in the high temperature limit.         


\section{Quark Spectrum}

The present model can be simplified further by putting the left and right handed quarks on a discrete grid spanned by the 
spatial variable ${\bf x}=1,2,...N_x$ and choosing the auxiliary field $P$ to be a constant in space and time.  
In frequency space this sets the restriction that only certain Matsubara modes can interact ($n=m, k=l$).

\begin{equation}
\int_0^\beta d\tau \biglb( (\psi^\dagger\psi)^2+(\psi^\dagger i\gamma_5 \psi)^2\bigrb) 
= 4\beta \sum_{n,m,k,l} \delta_{n,m}\delta_{k,l}\psi^\dagger_{Rn}\psi_{Rm}\psi^\dagger_{Lk}\psi_{Ll}.
\end{equation}
One can now bosonize quark pairs of opposite chirality in eq.~(\ref{eq:3}) using the auxiliary 
matrix $W^{x,y}_{n,m}=\psi^x_{Rm}\psi^{\dagger y}_{Ln}$ where the upper indices refer to three-space 
and the lower indices to frequency space resulting in the following Lagrangian

\begin{equation}
\mathcal{L}_{4}=\psi^\dagger(\Omega\gamma_4+im+i\mu\gamma_4-A_4\gamma_4)\psi+\tilde{\Sigma} N_C 
\text{Tr}_{x,n}(WW^\dagger)+\psi^\dagger_R W \psi_L + \psi^\dagger_L W^\dagger \psi_R+U(\phi[A],\phi^*[A],T)
\end{equation}
where $\Omega=\omega_n \openone_n\otimes \openone_x \otimes \openone_C$.  The four-Fermi interaction causes 
the quarks to interact as if they were moving in a random Gaussian potential provided by the auxiliary fields \cite{Janik:1998td}
\begin{eqnarray}
P=-2ig^2\langle\psi^\dagger_R\psi_L\rangle \nonumber\\
P^\dagger=-2ig^2\langle\psi^\dagger_L\psi_R\rangle
\end{eqnarray}

Note that in the above Lagrangian we have defined $\Tilde{\Sigma}=\beta\Sigma=V_4/(2g^2)$.  This was done in order to make a connection with the {\em standard} RMM used in the literature.  We will first look at this {\em standard} model in the thermodynamic limit where ${\bf n}=V_4/N=const$. Then we will relax this assumption and include the temperature dependence in the action ({\em i.e.} ${\bf n}=V_3/N=const.$).  Note that the correct temperature weight is required in order for the matrix model to reproduce the mean field results with the lowest two matsubaras.  

We first set $\phi[A]$ to a fixed value so the potential $U$ will not affect the dynamics.  The resulting form of the partition        function allows for the investigation of the quark spectrum in the presence of the background gauge field.
\begin{equation}
Z=\int \mathcal{D}[W] e^{-N\Tilde{\Sigma}\text{Tr}WW^\dagger}\prod_{f=1}^{N_f}\text{det}\left( \begin{array}{cc}m & iW+i\omega_n+\mu-iA^4 \\ iW^\dagger+i\omega_n+\mu-iA^4 & m \end{array} \right).\label{eq:rmm}
\end{equation}
In the above model we have set the Dyson index to two corresponding to $N_c = 3$ and 
the matrix elements correspond to the chiral unitary ensamble ($\chi$GUE). Each matrix $W$ has 
$N = 2N_\omega N_{\bf x} N_c$ entries whose interaction matrix elements $W_{ij}$ are drawn from a gaussian 
distribution having variance $\tilde{\Sigma}=1$.

Without the inclusion of the $A_4$ term in the above random matrix model (RMM), the above partition function 
is the chiral random matrix model of \cite{Papp:1998uz}.  The addition of the background gauge field $A_4$ 
serves as an imaginary color chemical potential.  A similar model was considered in \cite{Wettig:1995fg} where 
a non-random component was added to the Dirac matrix in order to simulate the formation of instanton--anti-instanton pairs. 

We now examine the above matrix model for $N_f=1$ and $m=\mu=0$ and restrict the frequency space to the two lowest Matsubara modes.  
In this case $\phi = \overline{\phi}$ and the eigenvalues will be real.  The matrix model is composed of a random part 
$\bf{R}$ and a deterministic part $\bf{D}$.  The model can be re-written as 

\begin{equation}
Z=\int \mathcal{D}[R] e^{-N\Tilde{\Sigma}\text{Tr}_{\text{{\bf x},n,N}}RR^\dagger}\text{det}_{\text{{\bf x},n,N}}{\bf Q}
\end{equation}
where
\begin{equation}
{\bf Q} = \left( \begin{array}{cc}0 & {\bf D} \\ {\bf D}  & 0 \end{array} \right) + \left( \begin{array}{cc}0 & {\bf R} \\ {\bf R}^\dagger & 0 \end{array} \right)\label{Dirac}
\end{equation}
\noindent and $D=\openone_x\otimes\text{diag}(\pi T + \nu, \pi T, \pi T-\nu,-\pi T + \nu, -\pi T, -\pi T - \nu)$with $\nu=T\arccos(\frac{3\phi-1}{2})$.

In the mean-field approximation, the resolvent for the RMM follows readily from the use of Blue's
functions (B) which are the inverse of Green's functions or resolvents (G), ie $B(G)=G(B)=z$
\cite{BLUE}. The Blue's function for the random part is $B_R=z+1/z$, while that of the deterministic
part is

\begin{equation}
B_D(z)=\frac{1}{6}\sum_{n=1}^6\frac{1}{z-\bf{D}_n}
\end{equation}
where ${\bf D}_n$ represents the $n^{th}$ diagonal entry of the matrix ${\bf D}$.  The Blue's function 
for the RMM follows from the self-energy addition rule $B_{R+D}=B_R+B_D-1/z$. The Green's function or
resolvent for the RMM follows from the inverse rule $B_{R+D}(G)=z$,

\begin{equation}
G(z)+\frac 16 \sum_{n=1}^6\frac 1{G(z)-\bf{D}_n}=z
\end{equation}
which is a seventh order algebraic equation for $G(z)$,

\begin{equation}
G^7+a_6 G^6+a_5G^5++a_4G^4+a_3G^3+a_2G^2+a_1G+a_0=0
\label{E1}
\end{equation}
with
\begin{eqnarray}
a_6&=&-6z \nonumber\\
a_5&=&1-3\pi^2T^2-2\nu^2+15z^2 \nonumber\\
a_4&=&z (-5 + 12 \pi^2 T^2 + 8 \nu^2 - 20 z^2) \nonumber\\
a_3&=&3 \pi^4 T^4 + \nu^4 + 5 z^2 (2 + 3 z^2) - 2 \pi^2 T^2 (1 + 9 z^2) - 4/3 \nu^2 (1 + 9 z^2) \nonumber\\
a_2&=&-2 z (3 \pi^4 T^4 + \nu^4 - 3 \pi^2 T^2 (1 + 2 z^2) - 2 \nu^2 (1 + 2 z^2) + z^2 (5 + 3 z^2)) \nonumber\\
a_1&=&-\pi^6 T^6 + \nu^4 (1/3 + z^2) - 2 \nu^2 z^2 (2 + z^2) + z^4 (5 + z^2)\nonumber\\
   &\,& + \pi^4 T^4 (1 + 2 \nu^2 + 3 z^2) - \pi^2 T^2 (\nu^4 + 3 z^2 (2 + z^2))\nonumber\\
a_0&=&-z/3 (3 \pi^4 T^4 + \nu^4 - 6 \pi^2 T^2 z^2 - 4 \nu^2 z^2 + 3 z^4)\nonumber\\
\label{E2}
\end{eqnarray} 
The spectral density follows from $G(z)$ through its discontinuity along the real axis
\begin{eqnarray}
\rho(\lambda)=-\frac 1\pi\,{\lim_{\epsilon\to0}}\,{\rm Im}G(\lambda +i\epsilon)
\label{E3}
\end{eqnarray}
The algebraic solutions of (\ref{E1}) leading to (\ref{E3}) will be discussed elsewhere.

Now we discuss the case when the explicit temperature dependence is included in the Gaussian weight.  
The model is now written as
\begin{equation}
Z=\int \mathcal{D}[R] e^{-N\beta{\Sigma}\text{Tr}_{\text{{\bf x},n,N}}RR^\dagger}\text{det}_{\text{{\bf x},n,N}}{\bf Q}
\end{equation}
where $\Sigma=V_3/(2g^2)$.  The prior result for the resolvent corresponding to the {\em standard} matrix model can be re-used for the above matrix model via the re-scaling of $T\to\sqrt{T}$ and                $\rho(\lambda)\to \sqrt{\beta}\rho(\lambda\sqrt{\beta})$.

A key {\it difference} with the {\em standard} chiral RMM~\cite{rmm} is the fact that
the Gaussian R-weight factor has an explicit $\beta=1/T$ which causes the Gaussian weight to weaken as we cross 
the transition temperature from above going to low temperatures.  It is this mechanicsm which caused the chiral condenstate to vanish at zero temperature in the mean field analysis of the two lowest matsubara modes.  Our RMM is suited for studying spectra near $T_c$ as 
it embodies the extra suppression in $T$ encoded in the time-integration.

\begin{figure}[hbtp]
  \vspace{1pt}
  \centerline{\hbox{ \hspace{0.0in} 
\includegraphics[scale=.35]{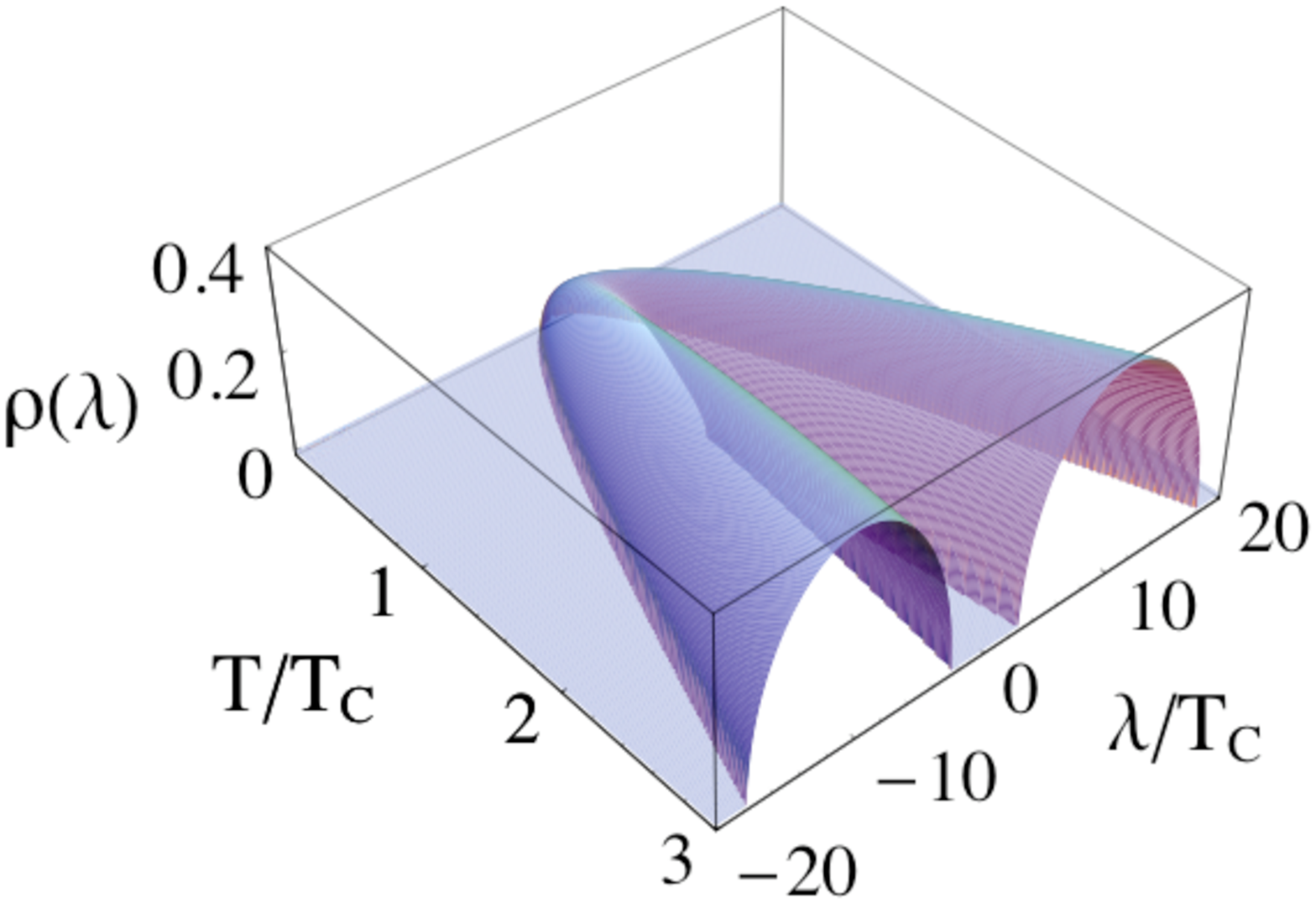}
\includegraphics[scale=.35]{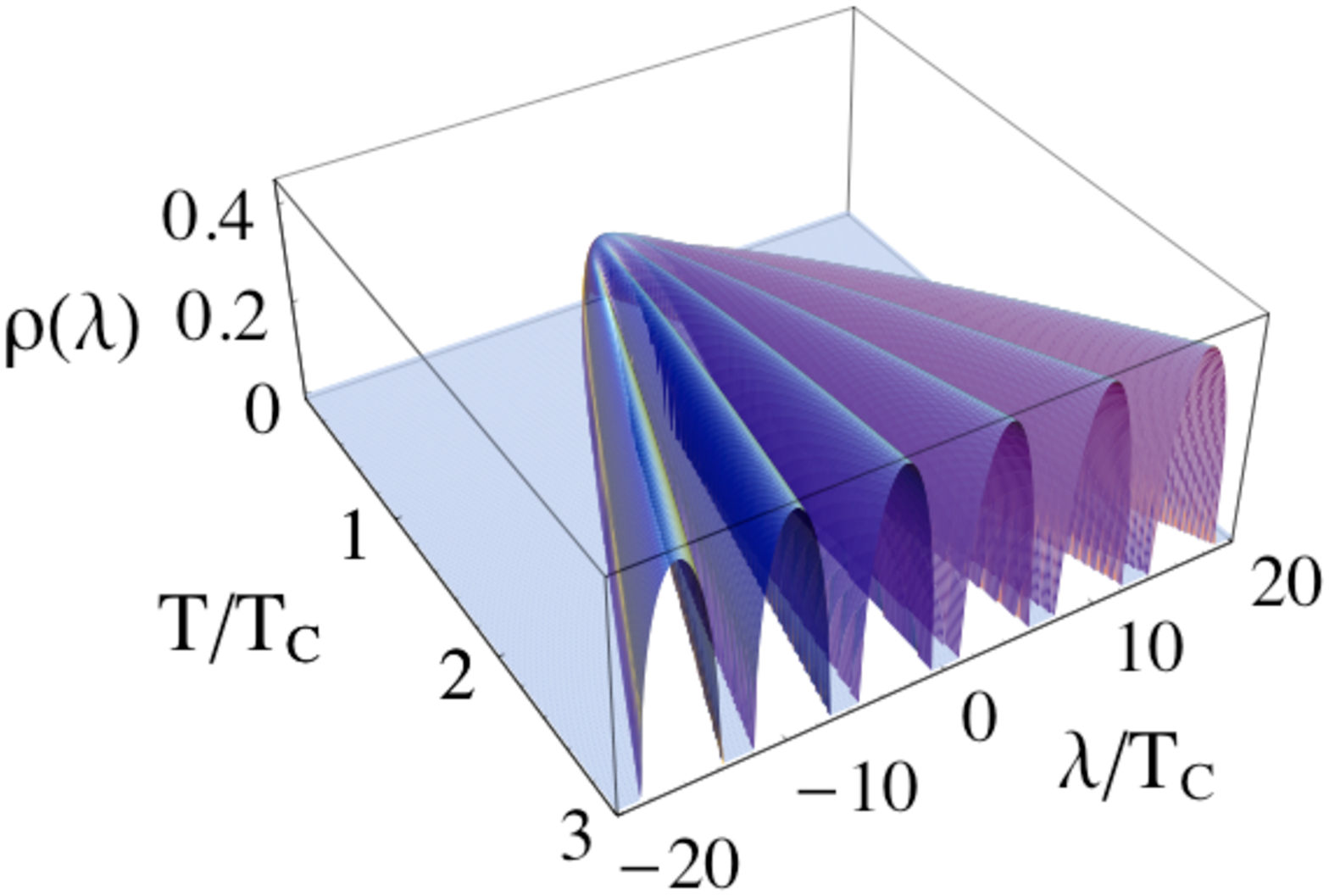}
    }
  }
   \vspace{1pt}
\caption{Spectral function as a function of $T$ for $\phi=1$ (left) and $\phi=0$ (right).}
 \label{fig:spec}
\end{figure}

Before we show the spectra we must discuss what values of the background potential to use.  The most physical choice of $\phi$ to take when computing spectra would be along 
the trajectory shown in figure~\ref{fig1}.  Instead, however, we simply choose to 
show spectra using $\phi=0,1$ in order to demonstrate the maximum effect of the 
Polyakov line. The spectra are shown in figure~\ref{fig:spec} as
functions of temperature for $\phi=1$ (left figure) and $\phi=0$ (right figure). 
The inclusion of the Polyakov line causes the 
spectra to split into separate domains as seen in the right figure.  

Of course, all of the above values of $T$ and $\phi$ are not realized in nature.  For the case of zero chemical potential 
(which is what is examined in this work) the value of $A_4=\text{diag}(\nu,0,-\nu)$ where $\nu=T\arccos(\frac{3\phi-1}{2})$.  
At low temperatures $\phi\approx0$ and $\nu\propto T$.  At higher temperatures $\phi\to1$ and $\nu\to0$ as dictated by the 
$\arccos$ dependence.  The maximum $\nu$ occurs a little above $T_C$.  We therefore expect to see the strongest changes in 
the eigenvalues near $T_C$.

\begin{figure}[hbtp]
  \vspace{9pt}
  \centerline{\hbox{ \hspace{0.0in} 
\includegraphics[scale=.7]{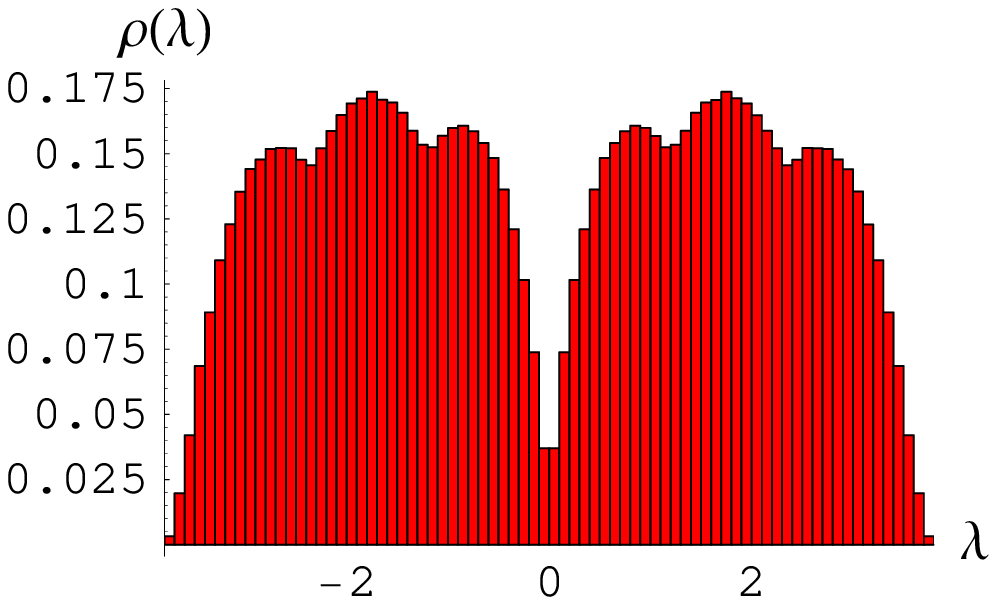}
    }
  }
  \vspace{9pt}
  \centerline{\hbox{ \hspace{0.0in}
\includegraphics[scale=.7]{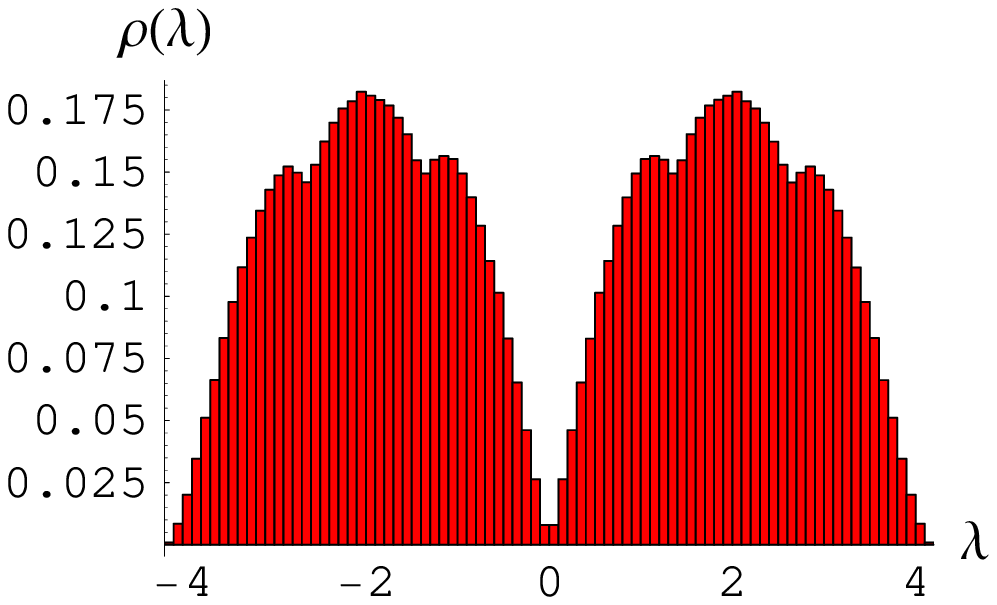}
    }
  }
 \vspace{9pt}
  \centerline{\hbox{ \hspace{0.0in} 
\includegraphics[scale=.7]{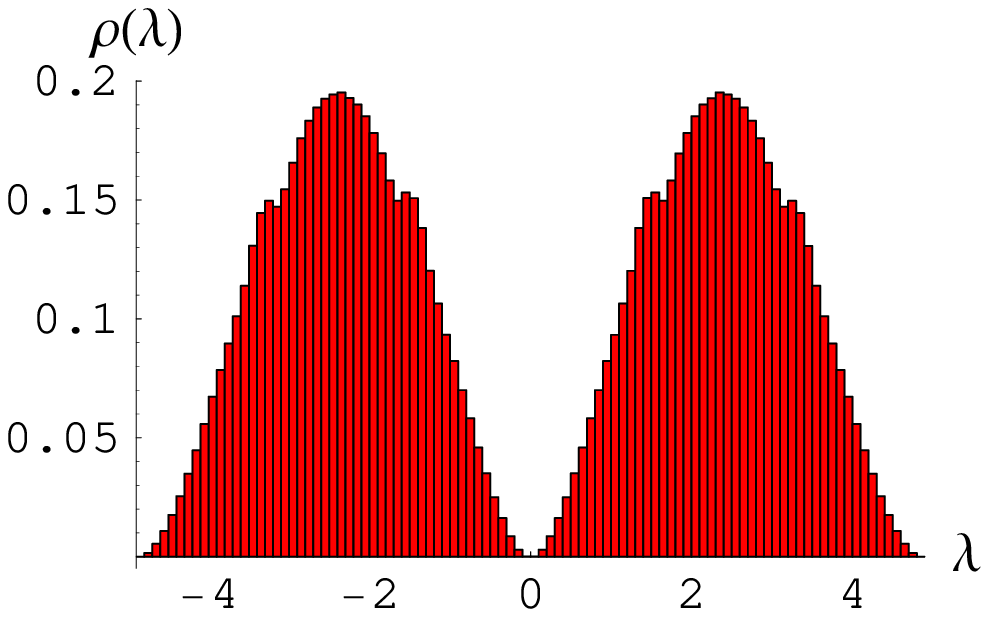}
    }
  }
  \vspace{9pt}
\caption{Spectral density of Dirac eigenvalues generated from an ensemble of 2000 $240\times240$ matrices 
at $T=0.75 T_C \text{ (top) }, T=T_C \text{ (middle) and } T=1.5 T_C \text{ (bottom)}$.  The value of 
$\phi$ used in each matrix was sampled from the distribution given by $U(\phi,T)$.}
  \label{fig:avgm}
\end{figure}

Finally, we compute the eigenvalue density by taking an ensemble average over $\phi$, which is the analogue of integrating
over the large gauge configurations across the transition temperature.  This is done at three temperatures, $0.75 T_c, T_c 
\text{ and } 1.5 T_c$.  The distribution function for $\phi$ is found from the potential $U(\phi)$.  At $T_c$ the probability 
distribution function for $\phi$ is peaked at $\phi=0$ and decreases monotonically to zero at $\phi=1$.  Below $T_c$ there 
is even more strength at lower values of $\phi$.  Above $T_c$ the distribution develops a peak at a finite value of $\phi$.  
The result of this ensemble averaging is shown in fig~\ref{fig:avgm}.

We find that the inclusion of a background gauge field brings about qualitative differences in the macroscopic spectral 
density in comparison to the standard chiral random matrix model.  There are oscillations present in the bulk of the 
spectrum that increase with temperature.  These oscillations are most likely above the Thouless energy and therefore 
in the diffusive regime of the lattice data.  However, there are also qualitative changes in the low energy eigenvalues.  
In the standard chiral random matrix model the spectral density is a semi-circle even above $T_c$.  Therefore the slope 
of the spectrum at low energy is infinite.  The inclusion of the background field changes this picture.  At $T_c$ one 
can see that the slope of the eigenvalue distribution at $\lambda=0$ is finite.   

The Polyakov loop raises the critical temperature from the standard NJL model.  The mechanism for this is now clear by 
looking at the spectrum.  Near $T_c$ one makes an ensemble average over all values of $\phi$.  When the gauge field is 
present, the spectrum is modified and the eigenvalues are shuffled to lower values of $\lambda$.  Due to the weighting from 
$U(\phi,T)$, only a small portion of the eigenvalues is shifted to lower $\lambda$.  This gives rise to the slope seen 
in the fully integrated spectrum.  These additional low-lying eigenvalues therefore shift the critical temperature to 
higher values due to the Banks-Casher relation.

\section{Conclusions}

We have constructed a simplified version of the PNJL model in 0+1 dimensions that embodies the
essentials of the model in 3+1 dimensions. The bulk pressure and susceptibilities across the
transition temperature are shown to follow from the interplay between the lowest Matsubara mode
and the Polyakov line. Many features of the quark number susceptibilities in the flavor symmetric
case are analogous to the ones observed in full fledged lattice simulations, providing simple
insights to the dynamics at works in QCD. The flavor asymmetric results both at finite temperature
and chemical potential will be discussed in a forthcoming paper.

The PNJL model in 0+1 dimensions yields a simple chiral RMM with a Polyakov line. At finite 
temperature, the presence of the Polyakov line causes the macroscopic Dirac spectrum to be
multihumped across the transition temperature, which is a consequence of the Z(3) symmetry
breaking at high temperature
in QCD. The oscillations are smeared but not eliminated by averaging over the Z(3) vacuua,
a feature that should be accessible to lattice simulations in the diffusive or chiral regime.
The effects of Z(3) breaking on the spectra at finite chemical potential will be discussed elsewhere.

\acknowledgments

This work was supported in part by US-DOE grants DE-FG02-88ER40388
and DE-FG03-97ER4014.

\end{document}